\documentclass[useAMS,usenatbib]{mn2e}

\usepackage{graphicx}
\usepackage{amsmath}
\usepackage{indentfirst}
\def\cm {$\rm cm^{-3} \,$}
\def\kms {$\rm km\,s^{-1} \,$}
\def\ergs {erg\,s$^{-1}$}
\def\ergcma {erg\,s$^{-1}$\,cm$^{-2}$\,\AA$^{-1}$}
\def\msunyr {M$_\odot$\,yr$^{-1}$}

\title[Gas kinematics and excitation of 3C\,33]{Gas rotation, shocks and outflow within the inner 3 kpc of the radio galaxy 3C\,33}

\author[Guilherme S. Couto et al.]{Guilherme S. Couto$^{1,2}$\thanks{E-mail:gcouto@if.ufrgs.br}, Thaisa Storchi-Bergmann$^{2}$, Allan Schnorr-M\"uller$^{2}$\\
$^{1}$Universidade Federal de Santa Catarina, Departamento de F\'isica-CFM, CP 476, 88040-900, Florian\'opolis, SC, Brazil\\
$^{2}$Universidade Federal do Rio Grande do Sul, IF, CP 15051, 91501-970, Porto Alegre, RS, Brazil\\}

\begin{document}

\date{Accepted 2017 April 19. Received 2017 April 7; in original form 2017 February 9}

\pagerange{\pageref{firstpage}--\pageref{lastpage}} \pubyear{}

\maketitle

\label{firstpage}

\begin{abstract}

We present optical integral field spectroscopy -- obtained with the Gemini Multi-Object Spectrograph -- of the inner $4.0 \times 5.8\,$kpc$^2$ of the narrow line radio galaxy 3C\,33 at a spatial resolution of $0.58\,$kpc. The gas emission shows three brightest structures: a strong knot of nuclear emission and two other knots at $\approx 1.4$\,kpc south-west and north-east of the nucleus along the ionization axis. We detect two kinematic components in the emission lines profiles, with a ``broader component'' (with velocity dispersion $\sigma \ge 150$\,\kms) being dominant within a $\sim 1\,$kpc wide strip (``the nuclear strip'') running from the south-east to the north-west, perpendicular to the radio jet, and a narrower component ($\sigma \le 100$\,\kms) dominating elsewhere. Centroid velocity maps reveal a rotation pattern with velocity amplitudes reaching $\sim \pm 350\,$\kms in the region dominated by the narrow component, while residual blueshifts and redshifts relative to rotation are observed in the nuclear strip, where we also observe the highest values of the [N {\sc ii}]/H$\alpha$, [S {\sc ii}]/H$\alpha$ and [O {\sc i}]/H$\alpha$ line ratios, and an increase of the gas temperature ($\sim 18000\,$K), velocity dispersion and electron density ($\sim 500\,$\cm). We interpret these residuals and increased line ratios as due to a lateral expansion of the ambient gas in the nuclear strip due to shocks produced by the passage of the radio jet. The effect of this expansion in the surrounding medium is very small, as its estimated kinetic power represents only $2.6 - 3.0 \times 10^{-5}\,$ of the AGN bolometric luminosity. A possible signature of inflow is revealed by an increase in the [O {\sc i}]/H$\alpha$ ratio values and velocity dispersions in the shape of two spiral arms extending to $2.3\,$kpc north-east and south-west from the nucleus.

\end{abstract}


\section{Introduction}

\begin{figure*}
\centering
\includegraphics[width=\textwidth]{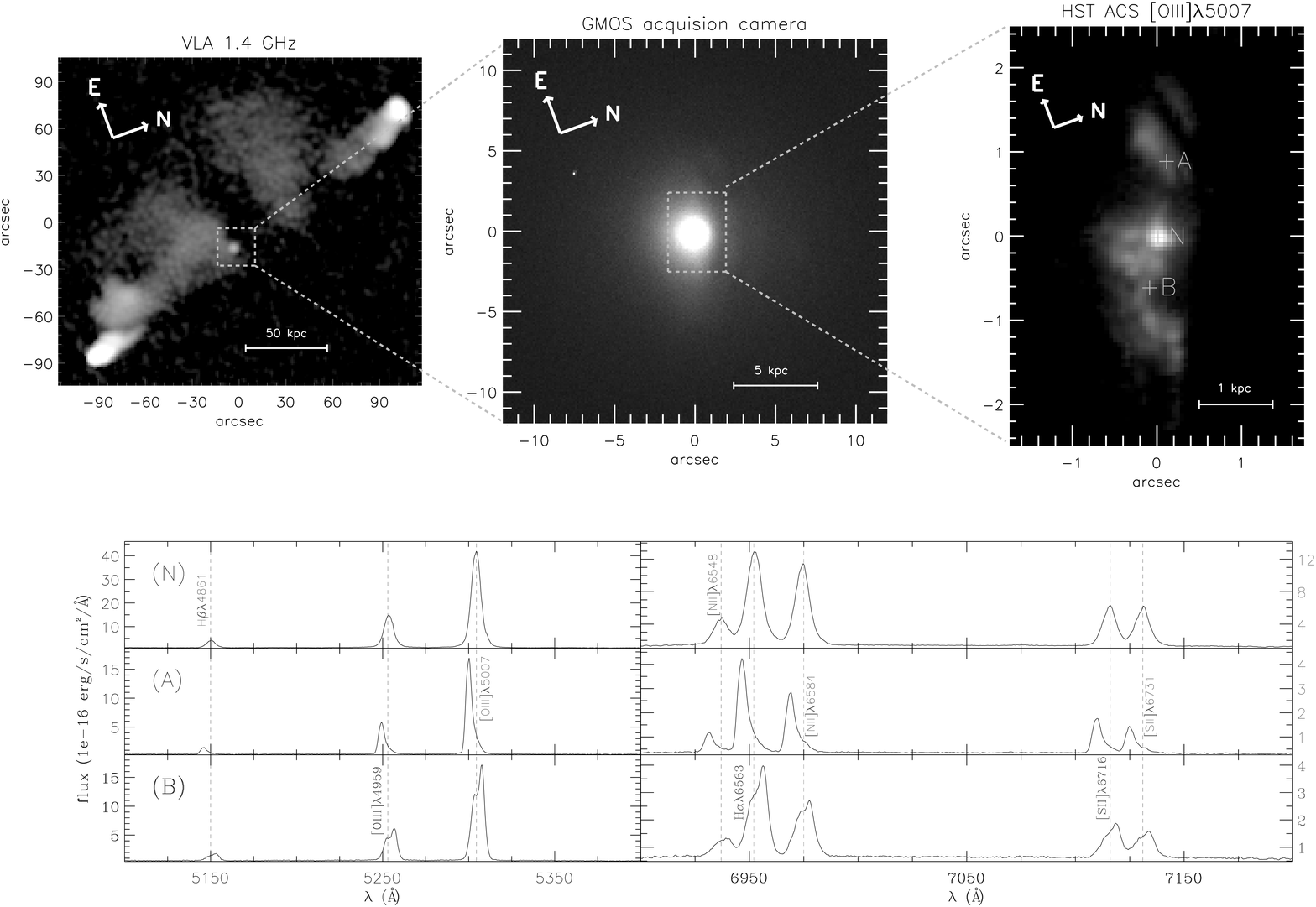}
\caption{Top left panel: 1.4 GHz VLA radio image of radio galaxy 3C\,33 and its lobes. top central panel: optical GMOS acquisition camera image of the galaxy. Top right panel: {\it HST-ACS} [O\,{\sc iii}]$\lambda$5007 narrow band image within the FoV of the GMOS IFU data. Bottom panels: Spectra extracted in positions N, A and B displayed in the top right panel.}
\label{large}
\end{figure*}

The connection between active galactic nuclei (AGN) and the host galaxy evolution has been a topic of debate in the past few decades, with substantial evidence of a link between the two. The tight correlation between the masses of the central supermassive black holes powering the AGN and the properties of the bulges of the host galaxies \citep[e.g.,][]{greene06,kormendy13}, indicate that AGNs and their host galaxies coevolve in some manner, at least partially due to the feedback from the AGN. Theoretical studies indicate that the feedback from the AGN is a necessary ingredient in the evolution of galaxies, preventing them from becoming too massive \citep{dimatteo05,wagner11,fabian12}. Radio activity, in particular, influences the energetics and thermodynamics of the gas, generating gas ionization by shocks, aside from photoionization due to the AGN radiation \citep{best00,fragile04,groves04a}. Radio galaxies are interesting laboratories to study the effects of AGN feedback such as energy injection into the interstellar medium (ISM) and quenching of star formation.

Observational evidence of interaction between the AGN feedback and the ISM in radio galaxies has been claimed by several studies \citep{tremblay09,couto13,santoro14,roche16}. Integral Field Spectroscopy (IFS) provide a direct observation of the impact of the AGN on the ambient gas. IFS can also reveal possible feeding mechanisms of the AGN that promote gas inflow towards the nucleus. In this work, we present IFS study of the radio galaxy 3C\,33.

3C\,33 is a nearby ($z = 0.0597$, $1''$ corresponds to $1.15\,$kpc in the galaxy\footnote{We adopt cosmological parameters $H_0 = 70.5\,$\kms\,Mpc$^{-1}$, $\Omega_\Lambda = 0.73$ and $\Omega_M = 0.27$}) Fanaroff-Riley type II radio galaxy, with its radio emission exhibiting two extended lobes, with VLBI imaging showing two symmetric jets aligned with a large-scaled structure, which extend up to $\sim 120\,$kpc each \citep{leahy91,giovannini05}. A high inclination angle of $75^\circ - 80^\circ$ between the radio jet axis and the line of sight is indicated by the ratio between the fluxes of the jet and counterjet. The optical emission line spectrum of 3C 33 is typical of Seyfert 2 galaxies, with an [O\,{\sc iii}] luminosity of $L_{[\textrm{O\,III}]} \sim 1 \times 10^{42}\,$erg\,s$^{-1}$. \citet{tremblay09}, using {\it HST-ACS} data, showed that the [O\,{\sc iii}] and H$\alpha$+[N\,{\sc ii}] emission extends up to $\sim 2.5\,$kpc to the north-east and south-west from the nucleus in an ``integral sign'' shape. Optical gaseous kinematic studies from \citet{simkin79} and \citet{heckman85} indicate rotation around an axis approximately aligned with the radio jet axis (position angle PA$\sim 20^\circ$). 

This work presents a two-dimensional analysis of the kinematics and excitation of the gas within the inner $\sim 2\,$kpc of 3C\,33. The paper is organized as follows: in Sec. \ref{obs} we describe the observations and the reduction of the data; in Sec. \ref{res} we explain our emission line fitting strategy, along with the resulting gas excitation and kinematic maps we obtain from it; in Sec. \ref{dis} we discuss the results and present our interpretations to explain the physical processes taking place in 3C\,33 circumnuclear region; and finally in Sec. \ref{conc} we present our conclusions.

\section{Observations and Data Reduction}
\label{obs}

The observations were obtained with the Integral Field Unit of the Gemini Multi Object Spectrograph (GMOS-IFU) at the Gemini North Telescope on August 17, 2010 (Gemini project GN-2010B-Q-66), in one-slit mode. Eight individual exposures of 3C\,33 of 940\,s were obtained, with a spectral coverage of $\lambda4500-7300\,$\AA, centered at $\lambda5900\,$\AA. The B600+\_G5307 grating with the IFU-R mask was used. The GMOS-IFU, using one-slit mode, has a rectangular field-of-view (FoV) of $\approx 3\farcs5 \times 5\arcsec$, corresponding to $4.0 \mathrm{kpc} \times 5.8 \mathrm{kpc}$ at the galaxy. The spectral resolution is R$\sim3800$ at $\lambda 7000\,$\AA, derived from the full width half maximum (FWHM) of the CuAr emission lines. The seeing during the observations was $0\farcs5$, as measured from the FWHM of a spatial profile of the calibration standard star. This corresponds to a spatial resolution of $\approx 580$\,pc at the galaxy.

The data reduction was accomplished using tasks in the {\sc GEMINI.GMOS IRAF} package as well as generic {\sc IRAF}\footnote{{\sc IRAF} is distributed by the National Optical Astronomy Observatory, which is operated by the Association of Universities for Research in Astronomy (AURA) under a cooperative agreement with the National Science Foundation.} tasks. The reduction process comprised bias subtracion, flatfielding, trimming, wavelength calibration, sky subtraction, flux calibration, differential atmosferic dispersion, building of the data cubes at a sampling of $0\farcs1 \times 0\farcs1$, and finally combining the eight individual data cubes into a final data cube.

Bias subtraction from the science, flat and twilight raw frames were performed using the task {\sc gfreduce}. Since the bias used in the reduction is not overscan subtracted, we did not apply overscan subtraction to the raw frames. Flat lamps were used to obtain the flat-field response map for each associated science frame also using {\sc gfreduce}, in interactive mode. {\sc gswavelength} task were used to obtain the wavelength solution, using the arclamp exposures, applying a fifth-order Chevyshev polynomial to the profile. Since spectral dithering were performed in the observations, with half exposures being centered at $\lambda5850\,$\AA\, and the other half at $\lambda5950\,$\AA, we applied wavelength calibration to each set of data separately. RMS errors in the wavelength were typically $\sim 0.1\,$\AA, which represent $\sim 20\%$ of spectral sampling of $0.46\,$\AA\, per pixel. Flux calibration was performed using the {\sc gscalibrate} task, using a sensitivity function created from the standard star Feige 110, observed just prior to the flat and science exposures, which derived from the {\sc gsstandard} task. Finally, the cube creation for each science frame was performed using {\sc gfcube}, where we also performed the atmosferic dispersion setting the keyword {\sc fl\_atmd} on, and we combined the scince cubes into a final cube using {\sc imcombine}, correcting for offsets. 

\section{Results}
\label{res}

\begin{figure}
\includegraphics[width=0.5\textwidth]{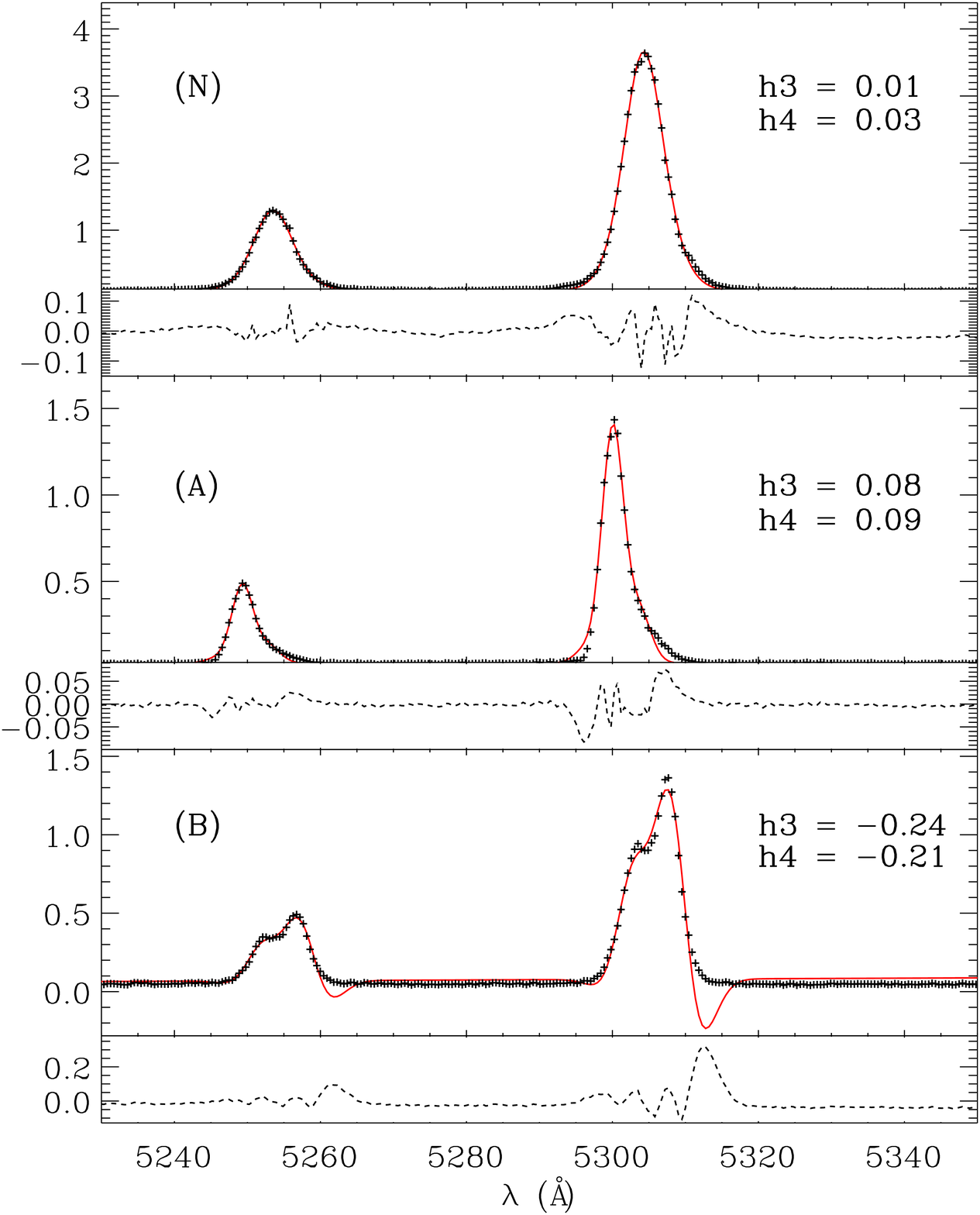}
\caption{Fit of Gauss-Hermite polynomials to the [O\,{\sc iii}]$\lambda\lambda$4959,5007 emission-line profiles, for the three positions shown in the top right panel of Fig. \ref{large}. Black crosses display spectrum data points and red lines represent the best model fitted to the profiles, with the resulting $h3$ and $h4$ parameters displayed in the upper right corner of each panel. Residuals are shown by dashed black lines. Flux units (y axis) are $10^{-16}\,$\ergcma.}
\label{ghfit}
\end{figure}

\begin{figure}
\includegraphics[width=0.5\textwidth]{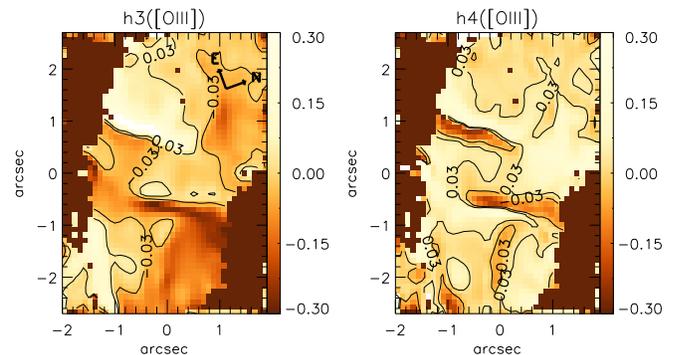}
\caption{Gauss-Hermite h3 and h4 parameter maps, obtained from the fit of the [O\,{\sc iii}]$\lambda\lambda$4959,5007 emission lines.}
\label{ghmap}
\end{figure}

\begin{figure*}
\centering
\includegraphics[width=\textwidth]{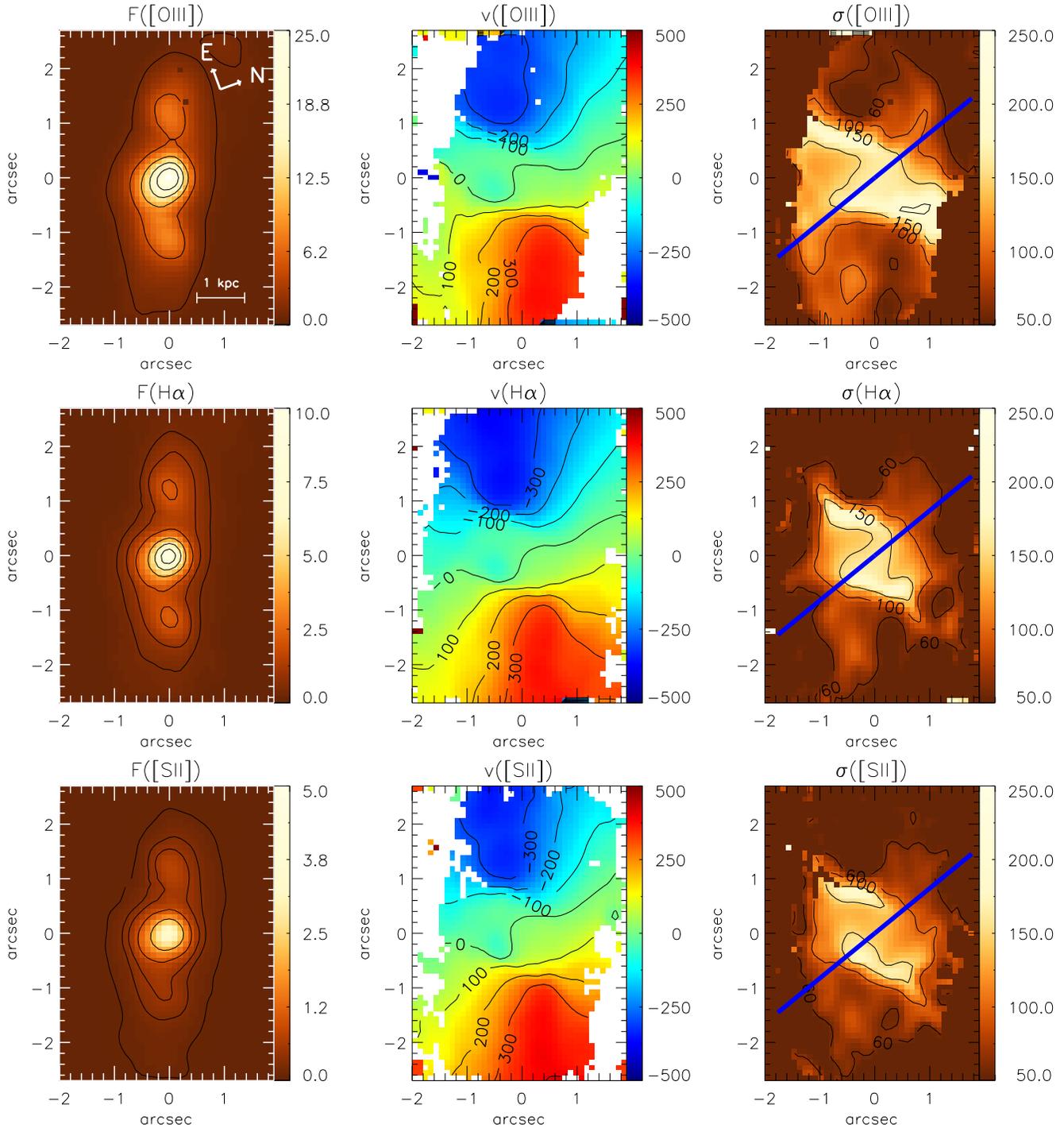}
\caption{Line flux, centroid velocity and velocity dispersion distributions resulting from the Gauss-Hermite fit of the [O\,{\sc iii}]$\lambda$5007, H$\alpha$ and [S\,{\sc ii}]$\lambda$6717 emission lines. The blue line shows the orientation of the radio jet axis. Flux units are $10^{-16}\,$erg\,s$^{-1}\,$cm$^{-2}\,$spaxel$^{-1}$. Centroid velocity and dispersion velocity units are \kms.}
\label{ghmaps}
\end{figure*}

\begin{figure*}
\centering
\includegraphics[width=\textwidth]{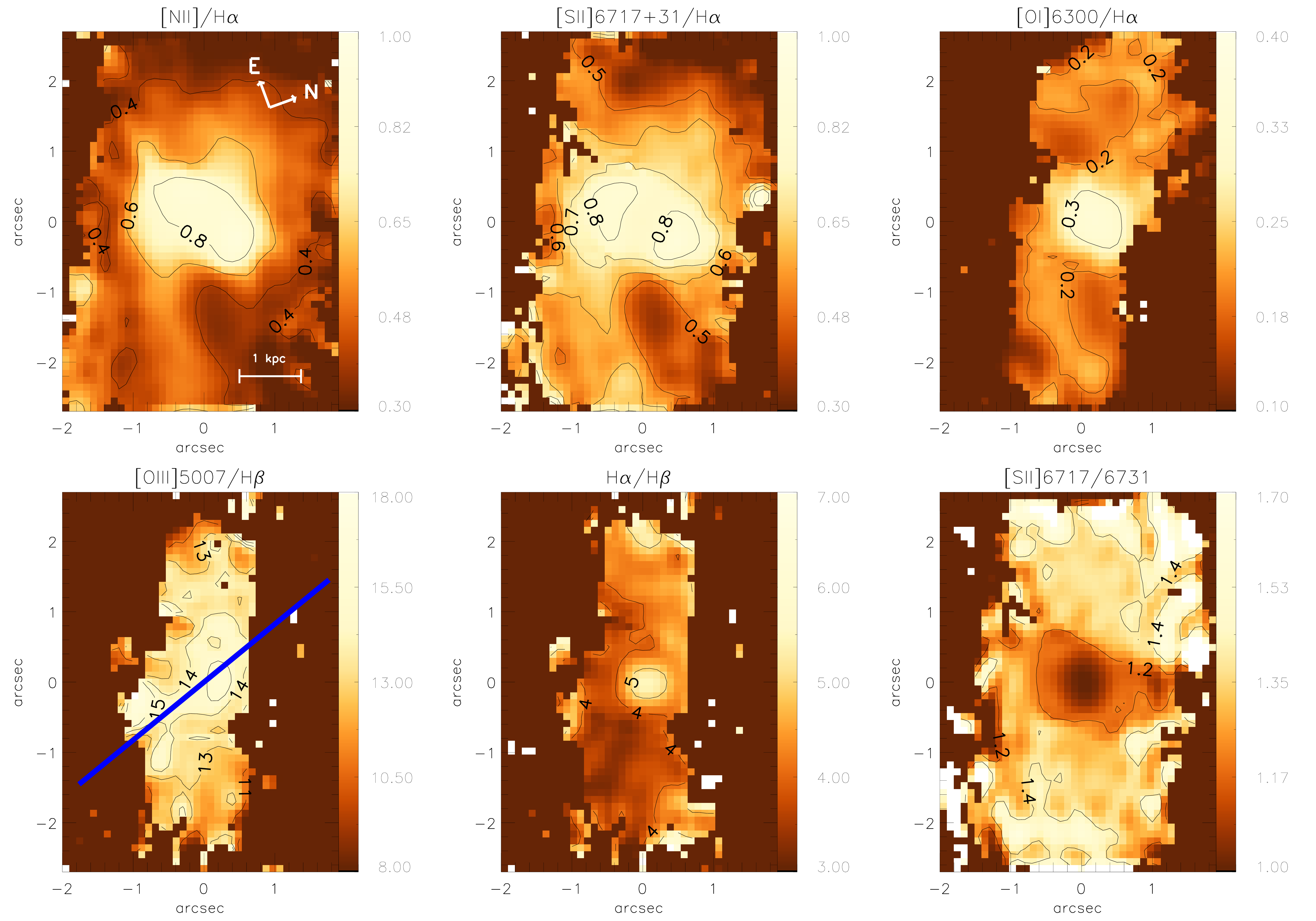}
\caption{Line ratio maps, obtained from the flux distributions derived from the Gauss-Hermite fit. The blue line displays the radio jet axis.}
\label{ghratios}
\end{figure*}

Fig. \ref{large} illustrates our IFU data and the FoV scale compared to other structures in 3C\,33. In the top left panel we display a 1.4 GHz Very Large Array (VLA) radio image, originally published by \citet{leahy91}, with a beam size of $4\arcsec$, covering a region of $\approx 250\arcsec \times 200\arcsec$. We have tilted the VLA image to the same orientation of our GMOS data. The image shows two lobes from the radio jet extending roughly to the north and south from the nucleus, along a position angle (PA) of $\approx -19.5$. Faint emission is observed between the two lobes, along with a hotspot in the galaxy core. The galaxy is displayed in the top central panel, in an image obtained with the GMOS acquisition camera ($\lambda$5620-6980, filter r\_G0303), covering a region of $10\arcsec \times 10 \arcsec$. Faint dark lanes are observed north from the nucleus, extending along the east-west direction, probably due to dust in a disk-like structure, suggesting that this is the near side of the galaxy. The top right panel shows a {\it HST-ACS} narrow band image\footnote{obtained from http://archive.stsci.edu, proposal ID 10882.} using the F551N ramp filter in the Wide Field Channel (WFC), with a central wavelength of $\lambda$5302, within a region of $\approx 3\farcs5 \times 5\arcsec$, which is the FoV of our GMOS IFU data. This [O\,{\sc iii}]$\lambda$5007 restframe-centred image displays its peak emission in the nucleus (position N), with extended emission resembling spiral arms, where we identify positions A and B, from which we extracted sample spectra.

The bottom panels display the spectra extracted in the positions N (nucleus, adopted as the location of the continuum peak flux), A (located $\approx 1\arcsec$ north-east from the nucleus) and B (located $\approx 0\farcs6$ south-west from the nucleus) shown in the top right panel, from the GMOS IFU datacube. Each spectrum has a spatial aperture of $0\farcs3$ radius, similar to our spatial resolution. We show in Fig. \ref{large} the profiles of the main emission lines: H$\beta \lambda 4861$, [O\,{\sc iii}]$\lambda\lambda$4959,5007, [N\,{\sc ii}]$\lambda\lambda$6548,84, H$\alpha \lambda 6564$ and [S\,{\sc ii}]$\lambda\lambda$6717,31. The [O\,{\sc i}]$\lambda\lambda$6300,34 emission lines are also strong, while He\,{\sc ii}$\lambda$4686, H$\gamma \lambda 4340$ and [O\,{\sc iii}]$\lambda$4363 are weaker. The line profiles display strong red and blue wings at positions A and B, respectively. In fact these wings are due to other kinematic components that will be discussed in the next sections. 

\subsection{One-component fit with Gauss-Hermite Polynomials}
\label{gaussh}

The gaseous centroid velocities, velocity dispersions and integrated fluxes were obtained by fitting Gauss-Hermite polynomials to the emission lines. The $h3$ and $h4$ Gauss-Hermite moments parametrize the deviations from a Gaussian profile, thus are good tracers of multiple emission-line components. $h3$ is related to the skewness of the profiles, and $h4$ to its kurtosis. In other words, $h3$ measures asymmetric deviations from a Gaussian profile, such as blue or red wings, and $h4$ quantify the peakiness of the profile, with $h4 > 0$ for a more peaked and $h4 < 0$ for a less peaked profile than a Gaussian curve. A Gaussian profile is obtained when $h3 = h4 = 0$. In order to reduce the number of free parameters in the fit, the following physically motivated constraints were imposed:

\begin{enumerate}
\item{different lines from the same ionic species have the same kinematic parameters. For example, the [S\,{\sc ii}]$\lambda\lambda$6717,31 emission lines have the same centroid velocity and velocity dispersion. This was also done for the Gauss-Hermite parameters $h3$ and $h4$;}
\item{the [N\,{\sc ii}]$\lambda\lambda$6548,84 emission lines have the same centroid velocity and velocity dispersion as H$\alpha$;}
\item{the [N\,{\sc ii}]$\lambda$6548 flux was fixed as 1/3 of the [N\,{\sc ii}]$\lambda$6584 flux, in accordance with nebular physics \citep{osterbrock06}. This was also done for the [O\,{\sc iii}]$\lambda\lambda$4959,5007 and [O\,{\sc i}]$\lambda\lambda$6300,34 emission lines.}
\end{enumerate}

In Fig. \ref{ghfit} we display the best Gauss-Hermite polynomial model fit to the [O\,{\sc iii}]$\lambda\lambda$4959,5007 emission lines, in the same positions marked in the top right panel of Fig. \ref{large}. We also display the Gauss-Hermite parameters $h3$ and $h4$ for each position. Noticeably Gauss-Hermite polynomials are not a very good model for the emission-line profiles of some regions, such as positions A and B. However, for others, it does fit the profile nicely, like in position N. The distribution of the $h3$ and $h4$ values for the [O\,{\sc iii}] emission lines are shown in Fig. \ref{ghmap}. Regions which present $|h3| > 0.03$ point to the presence of red (at locations mainly $\approx\,1''$ east from the nucleus) and blue wings (at locations mainly $\approx\,1''$ west from the nucleus). The $h4$ values reach high negative values in two narrow strips running approximately south-north, $\approx\,1''$ to the east and $\approx\,1''$ to the west of the nucleus, at the border of the regions with the highest positive and negative values of $h3$. Inspection of the profiles in these regions reveal that they actually present two kinematic components, like the spectrum of position B, to the west of the nucleus. We thus decided to perform also two Gaussian fits to regions with $|h3|$ or $|h4|$ larger than $0.03$. We discuss these fits further in the next subsection.

The measured integrated flux, centroid velocity and velocity dispersion maps for [O\,{\sc iii}]$\lambda$5007, H$\alpha$ and [S\,{\sc ii}]$\lambda$6717 emission lines, resulting from the Gauss-Hermite fits, are displayed in Fig. \ref{ghmaps}. Flux distributions are similar to that of the [O\,{\sc iii}] {\it HST} image, shown in the right panel of Fig. \ref{large}. These flux distributions display, besides a strong emission peak at the nucleus, two other regions of enhanced emission to the north-east and south-west of the nucleus, at the locations of the spiral arms seen in the {\it HST} image. The centroid velocity maps display a pattern almost identical to the known ``spider'' diagram, expected for rotation, with the kinematic major axis approximately aligned with the region of most extended emission in the flux maps. Blueshifts and redshifts are observed to the east and west of the nucleus, respectively, with high velocity amplitudes ($\sim \pm 350\,$\kms). There is a noticeable distortion in the rotating pattern in the inner arcsecond or so, where we conclude there is a second kinematic component (see discussion below). This region also presents higher velocity dispersions $\sigma$ reaching $\sim 170\,$\kms. This region of highest $\sigma$ has the shape of a band $\sim 2''$ wide centered at the nucleus and running approximately perpendicularly to the radio jet, represented by the blue line in Fig. \ref{ghmap}. Hereafter we will refer to this region as ``the nuclear strip''. Velocity dispersions are low outside these regions ($\sigma < 100\,$\kms), with the lowest values of $< 70\,$\kms displayed in the regions of highest centroid velocity amplitudes. Centroid velocities were obtained after the subtraction of the systemic velocity of $v_{sys} = 17801.8\,$\kms, resulting from the rotation model fitted to the H$\alpha\,$ velocity map, which we discuss further in Sec. \ref{rot_model}. Also, we corrected the velocity dispersion values for the instrumental broadening ($\sigma_{inst} = 44.3\,$\kms). 

Uncertainties in the Gauss-Hermite fits were estimated using Monte Carlo simulations of 100 iterations, in which Gaussian noise is added to the spectra. Expressing the flux uncertainty, $\epsilon_F$, as a fraction of the integrated flux, $F$, we find that the main emission lines present $\epsilon_F/F < 0.05$ throughout the whole FoV, except for the regions close to the borders of the datacube, where typical values of $\epsilon_F/F \sim 0.1$, but can reach up to $\epsilon_F/F \sim 0.3$ in weaker emission lines, like H$\beta$. The uncertainties of the centroid velocity and velocity dispersion measurements are similar, and display values of $\epsilon_v \sim \epsilon_\sigma < 10\,$\kms (but typically $\sim 2-3\,$\kms) in the nucleus and its surroundings. Close to the borders of the FoV, uncertainties in the velocity and velocity dispersions can reach $\sim 40\,$\kms. In Figs. \ref{ghmap} and \ref{ghmaps} (as in the other maps presented in this paper) we have masked out regions presenting uncertainty values higher than $\epsilon_F/F = 0.1$, $\epsilon_v = 10\,$\kms or $\epsilon_\sigma = 10\,$\kms.

Fig. \ref{ghratios} shows emission-line ratio maps obtained from the flux distributions. The highest values of [N\,{\sc ii}]/H$\alpha$, [S\,{\sc ii}]/H$\alpha$ and [O\,{\sc i}]/H$\alpha$ (value ranges of $0.6-0.9$, $0.7-0.85$ and $0.25-0.35$ for each of these three ratio distributions, respectively) are observed in the nucleus, and extended toward the nuclear strip, mainly in the former two ratio distributions. Intermediate values ($\sim 0.5$, $\sim 0.5$ and $\sim 0.2$ for [N\,{\sc ii}]/H$\alpha$, [S\,{\sc ii}]/H$\alpha$ and [O\,{\sc i}]/H$\alpha$ respectively) seems to extend towards the south-western border of our FoV. A counterpart of this extension is observed in the [O\,{\sc i}]/H$\alpha$ ratio towards north-east from the nucleus, but not in [N\,{\sc ii}]/H$\alpha$ and [S\,{\sc ii}]/H$\alpha$. The lowest ratio values ($\sim 0.3$, $\sim 0.35$ and $\sim 0.15$ for [N\,{\sc ii}]/H$\alpha$, [S\,{\sc ii}]/H$\alpha$ and [O\,{\sc i}]/H$\alpha$ respectively) seems to be located in the regions of highest velocity amplitudes, $\approx 1\farcs5$ east and west from the nucleus. The [O\,{\sc iii}]/H$\beta$ ratio is high throughout our FoV ([O\,{\sc iii}]/H$\beta > 10$), due to the strong emission of the [O\,{\sc iii}]$\lambda$5007 line. The highest values ($\sim 14-16$) are observed in the nucleus and extending in the same direction of the radio jet axis, indicated by the blue line in Fig. \ref{ghratios}. Line ratio values then drop with distance from the nucleus until they reach the lowest values ($\sim 11$) close to the borders of the FoV. The H$\alpha$/H$\beta$ ratio is high in the nucleus ($\sim 5.5$), with highest values $> 6$  observed towards the north-west. Values higher than H$\alpha$/H$\beta = 4$ are observed north from the nucleus, extending along the east-west direction, following the dark lane observed in the acquisition camera image (top central panel of Fig. \ref{large}). Lower values of H$\alpha$/H$\beta \sim 3.5$ are observed south of the nucleus. Finally, the last ratio map displayed in Fig. \ref{ghratios}, [S\,{\sc ii}]6717/31, shows the lowest values at the nucleus ($\sim 1$), and somewhat higher ($\sim 1.2$) in the nuclear strip. Higher ratios are observed outwards, reaching $\sim 1.5$ at the borders of the FoV.

\subsection{Two-components fit with Gaussian profiles}

\begin{figure}
\includegraphics[width=0.5\textwidth]{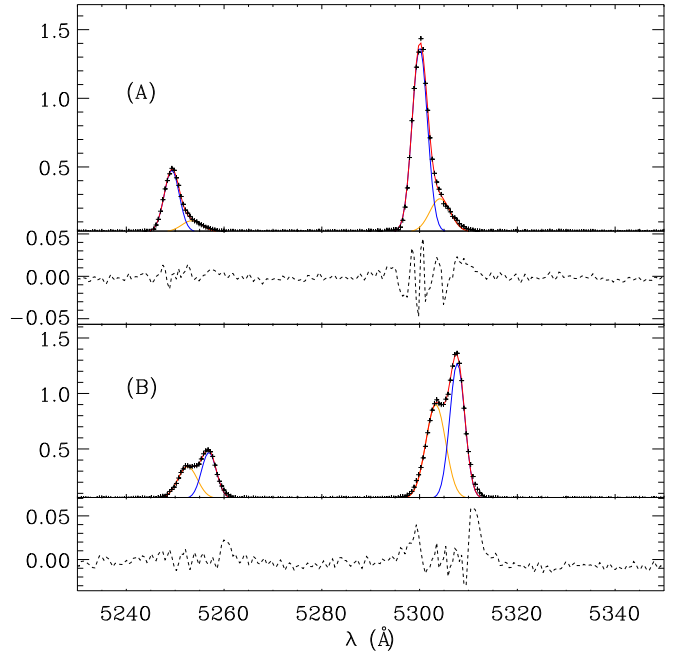}
\caption{Two Gaussians fit into the [O\,{\sc iii}]$\lambda\lambda$4959,5007 emission-line profiles, for the positions A and B shown in the top right panel of Fig. \ref{large}. Black crosses display spectrum data points and red lines represent the best model fitted to the profiles, with the narrow and broad component being represented by the blue and orange lines, respectively. Residuals are shown by dashed black lines. Flux units (y axis) are $10^{-16}\,$\ergcma.}
\label{2gfit}
\end{figure}

\begin{figure*}
\centering
\includegraphics[width=\textwidth]{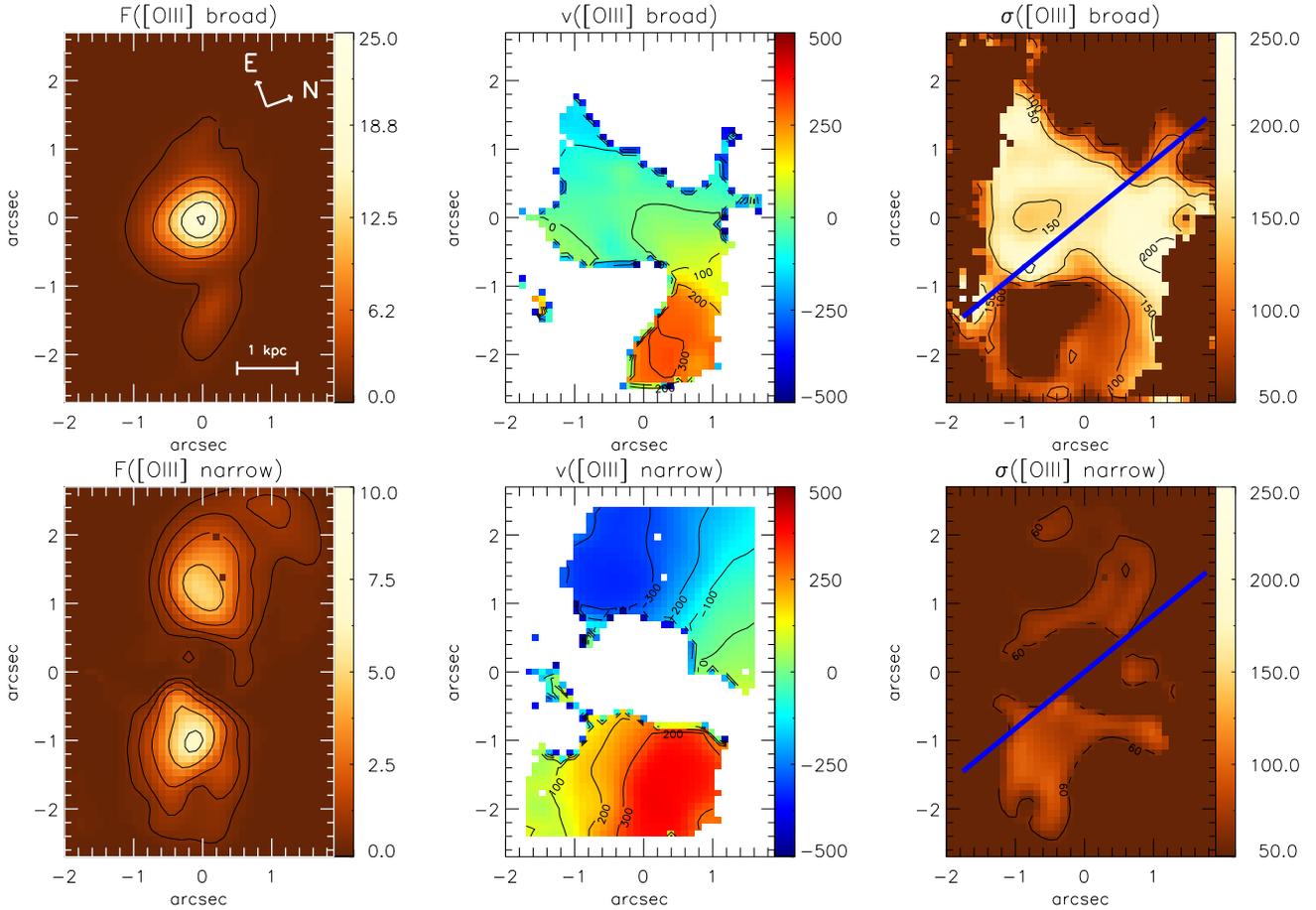}
\caption{Integrated flux, centroid velocity and velocity dispersion distributions resulted from the two Gaussians fit of the [O\,{\sc iii}]$\lambda$5007 line, for the broad and narrow component. The blue line displays the radio jet axis. Flux units are $10^{-16}\,$erg\,s$^{-1}\,$cm$^{-2}\,$spaxel$^{-1}$. Centroid velocity and dispersion velocity units are \kms.}
\label{2gmaps}
\end{figure*}

As mentioned in the previous section, for regions with $|h3|$ or $|h4|$ parameters (derived from the Gauss-Hermite polynomials fits) higher than $0.03$ we performed a two Gaussians fit in order to separate the two kinematic components of the emission-lines. We used the same physical constraints used for the Gauss-Hermite fits, with an additional one: that the ratio between the two components should be the same for lines of the same ionic species, like [O\,{\sc iii}]$\lambda\lambda$4959,5007. Since the two components are very blended in some regions, such as at and around the nucleus, they become ``degenerate'', and the fit is not well constrained. As the fits show that one component is broader than the other, we decided to identify the components as ``broad'' and ``narrow'' components. Examples of the fit of two components are shown in Fig. \ref{2gfit}, for the positions A and B. For regions with $h3$ and $h4$ close to $0$, we have performed a single Gaussian fit. We have then constructed separate flux and kinematic maps for the broad ($\sigma \ge 130\,$\kms) and narrow ($\sigma < 130\,$\kms) components.

We show in Fig. \ref{2gmaps} the result of the two Gaussian fits of the [O\,{\sc iii}]$\lambda$5007 profiles. We display the integrated flux, centroid velocity and velocity dispersion maps separately for the broad (top panels) and narrow (bottom panels) components. The flux distribution of the broad component is concentrated in the nucleus, with some extension towards the south-west and north-east within the nuclear strip. The emitting gas extending $\sim 1\farcs2$ south-west and north-east from the nucleus is traced by the narrow component. However, in other emission-line flux distributions, such as H$\alpha$, we derive more contribution of the narrow component close to the nucleus, but this is result of the blending of the two components in this region. The centroid velocity maps show that the rotation pattern is indeed traced by the narrow component, with the broader component presenting velocity values close to $0\,$\kms without a clear pattern. Some redshifts are observed in the ``tail'' structure of the broad component.

The $\sigma$ distributions show values of $\sim 90\,$\kms along the north-east and south-west of the nucleus for the narrow component, and $\sim 170\,$\kms for the broad component along the nuclear strip.

\subsection{Channel maps}

\begin{figure*}
\centering
\includegraphics[width=\textwidth]{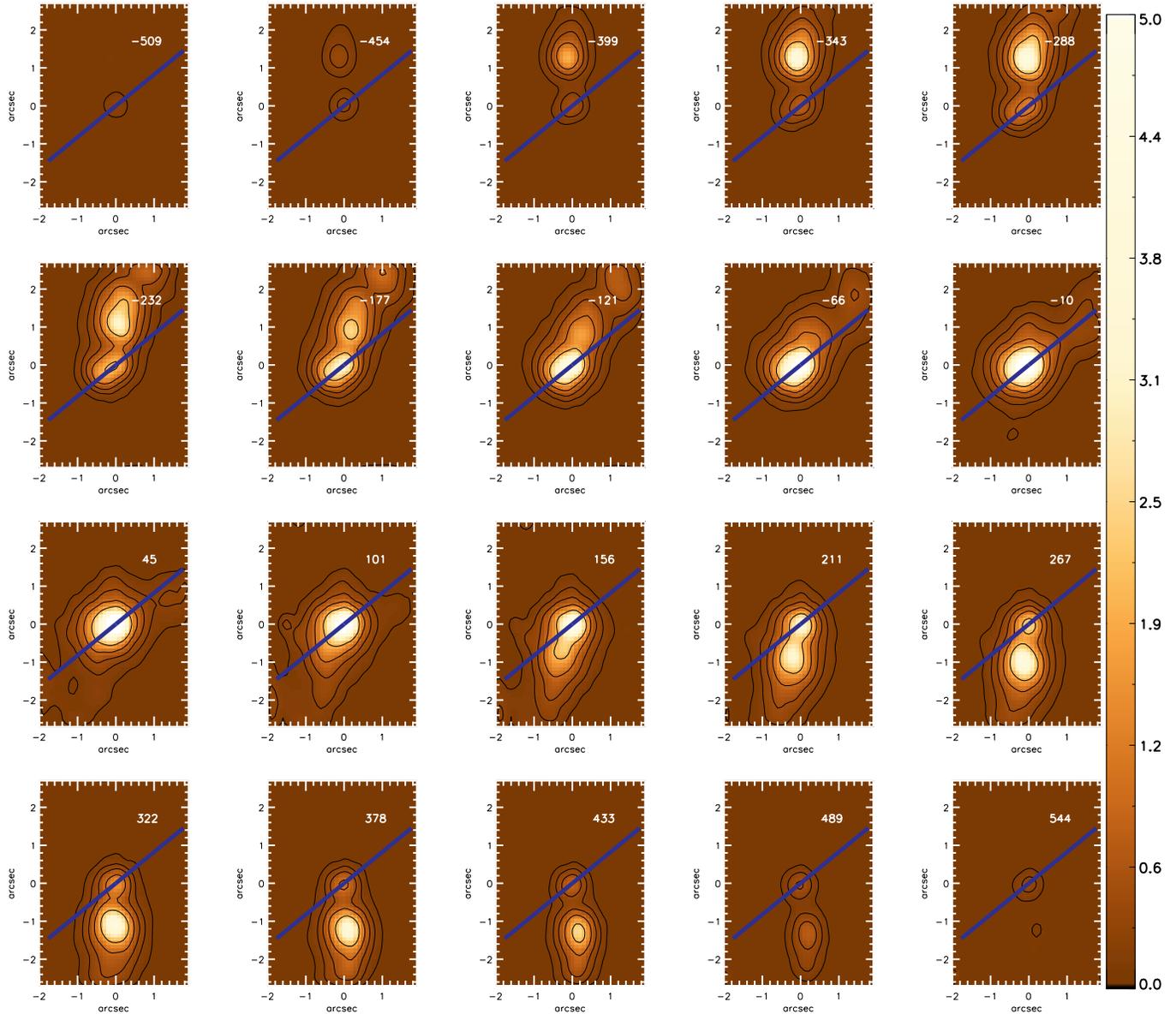}
\caption{Channel maps along the [O\,{\sc iii}]$\lambda$5007 emission-line profile, in order of increasing velocities shown in the top of each panel in units of \kms. Flux units are $10^{-16}\,$erg\,s$^{-1}\,$cm$^{-2}\,$spaxel$^{-1}$. The green dashed line displays the radio jet axis.}
\label{cmOIII}
\end{figure*}

\begin{figure*}
\centering
\includegraphics[width=\textwidth]{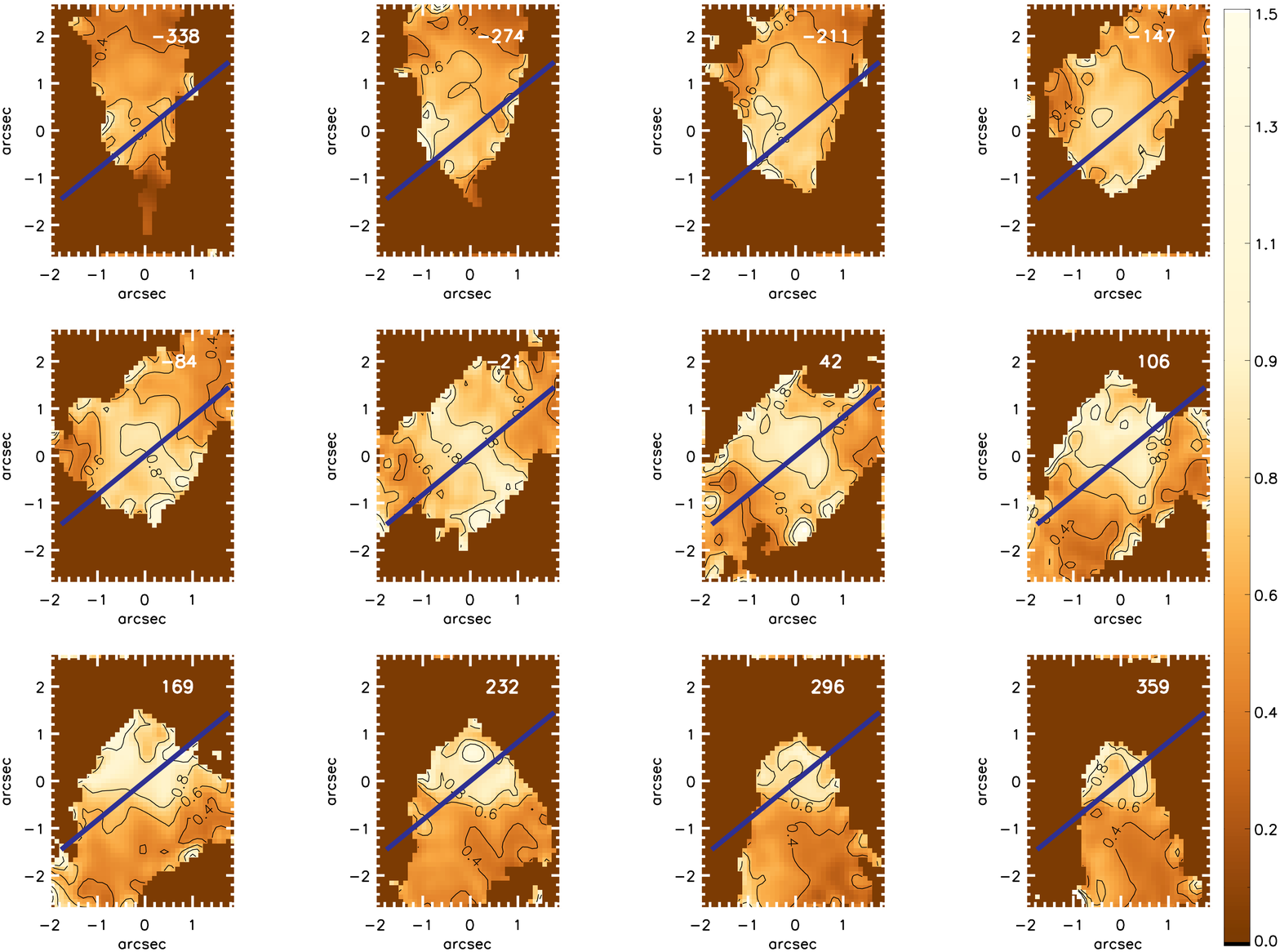}
\caption{Channel maps of the [N\,{\sc ii}]/H$\alpha$ emission-line ratio, in order of increasing velocities shown in the top of each panel in units of \kms. Velocity bins are averages of the [N\,{\sc ii}] and H$\alpha$ channels. The blue line displays the radio jet axis.}
\label{cmNIIHa}
\end{figure*}

\begin{figure*}
\centering
\includegraphics[width=\textwidth]{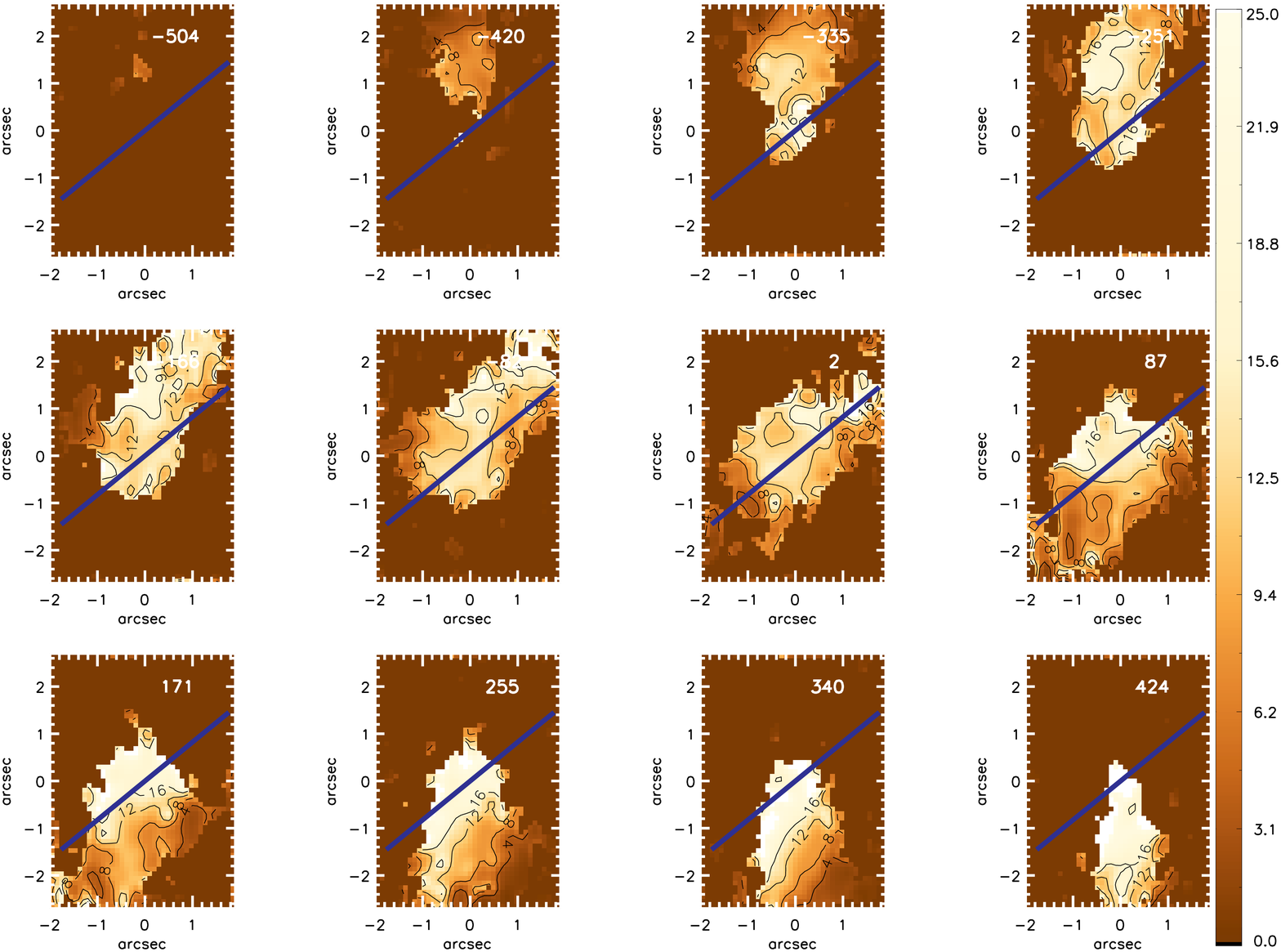}
\caption{Channel maps of the [O\,{\sc iii}]/H$\beta$ emission-line ratio, in order of increasing velocities shown in the top of each panel in units of \kms. Velocity bins are averages of the [O\,{\sc iii}] and H$\beta$ channels. The blue line displays the radio jet axis.}
\label{cmOIIIHb}
\end{figure*}

Fig. \ref{cmOIII} displays channel maps of the [O\,{\sc iii}]$\lambda$5007 emission-line profile. The maps were extracted in velocity bins of $\approx 55\,$\kms along the emission line. The channel maps reveal two main components: a nuclear component, barely resolved within the inner $0\farcs6$ ($\sim 700\,$pc) and an extra-nuclear extended component with blueshifted and redshifted velocities towards the north-east and south-west of the nucleus, respectively. The bulk of the extended emission has velocities of $\pm 300\,$\kms, corresponding to the extended emitting gas observed $\sim 1\farcs2$ north-east and south-west of the nucleus in the flux maps. However, extended emission is also observed in higher velocity bins, up to $\pm 450\,$\kms. At zero velocities (close to the systemic), we observe extended emission following the orientation of the radio jet.

\section{Discussion}
\label{dis}

\subsection{Gas excitation}

\subsubsection{Flux distributions}

The flux maps of the [O\,{\sc iii}]$\lambda$5007, H$\alpha$ and [O\,{\sc i}]$\lambda$6300 emission lines derived from the Gauss-Hermite fits (Fig. \ref{ghmaps}) show an elongated structure extending to $\approx 1.4\,$kpc north-east and south-west of the nucleus. In the two Gaussian fits, this structure is traced by the narrow component (Fig. \ref{2gmaps}). This narrow component dominates in regions farther than $\sim 700\,$pc from the nucleus, while the broad component dominates within the inner $\sim 700\,$pc but extending along the nuclear strip running from the south-east to the north-west, but at lower intensity levels when compared to the narrow component.

\subsubsection{Line-ratio maps}

Emission-line ratios typical of Seyferts are observed over the whole FoV as shown in Fig. \ref{ghratios}, in agreement with the strong [O\,{\sc iii}]$\lambda$5007 emission and high [O\,{\sc iii}]/H$\beta$ line ratios, which places all covered locations of our FoV well into the Seyfert region of BPT diagrams \citep{baldwin81,kewley06}. The highest ratios of [N\,{\sc ii}]/H$\alpha$, [S\,{\sc ii}]/H$\alpha$ and [O\,{\sc i}]/H$\alpha$ are observed along the nuclear strip. The increase in the latter line ratios can be attributed to the contribution of shocks \citep{allen08}. This hypothesis is supported by the fact that the radio jet runs approximately perpendicular to the nuclear strip.  We propose a scenario in which the radio jet is causing a lateral expansion of the ambient gas as it travels outwards from the nucleus, forming the observed strip of increased line ratios approximately perpendicularly to the radio jet. In the case of [O\,{\sc i}]/H$\alpha$, we also observe increased values up to ~2.5 kpc to the north-east and south-west in the shape of two spiral arms that seem to correlate with those seen in the {\it HST} [O\,{\sc iii}] image in Fig. \ref{large}. A possible scenario is that this structure is tracing inflowing gas. Along these arms, we also observe an increase in the velocity dispersion values (when compared to neighboring regions) which is consistent with the presence of shocks in these spiral arms, allowing the gas to move inwards to feed the SMBH.

We have used the channel maps in the emission lines to build also line-ratio channel maps of [N\,{\sc ii}]/H$\alpha$ and [O\,{\sc iii}]/H$\beta$, which allow us to investigate the excitation according to the kinematics of the gas. These are shown in Figs. \ref{cmNIIHa} and \ref{cmOIIIHb}. The highest [N\,{\sc ii}]/H$\alpha$ ratios are observed along the nuclear strip, most noticeable in the channels from $-84\,$\kms to $+106\,$\kms, thus close to zero velocity. In these channels, the highest ratios are observed in blueshift to the west and in redshift to the east. On the other hand, the highest [O\,{\sc iii}]/H$\beta$ ratios are observed along the radio jet axis. The radio jet, which extends to far greater distances than probed by our FoV, may have ``cleared a path'' in its way, pushing gas away in this direction and leading to an increased ionization parameter along this axis. 

We observe high-ionisaton emission from [Fe\,{\sc x}]$\lambda$6375, [Fe\,{\sc vii}]$\lambda$5721 and [Fe\,{\sc vii}]$\lambda$6087 at the nucleus with a suggested extension to the south. Higher signal to noise data would be necessary to confirm this extended emission, but if this is indeed the case, shock-driven outflows could be the responsible mechanism for the excitation, as previously observed in Seyfert galaxies \citep[e.g.][]{rodriguez-ardila06}, consistent with the scenario we propose where shocks are present near the nucleus.

\subsubsection{Electron density, O/H abundances, visual extinction and gas temperature}

\begin{figure*}
\centering
\includegraphics[width=0.8\textwidth]{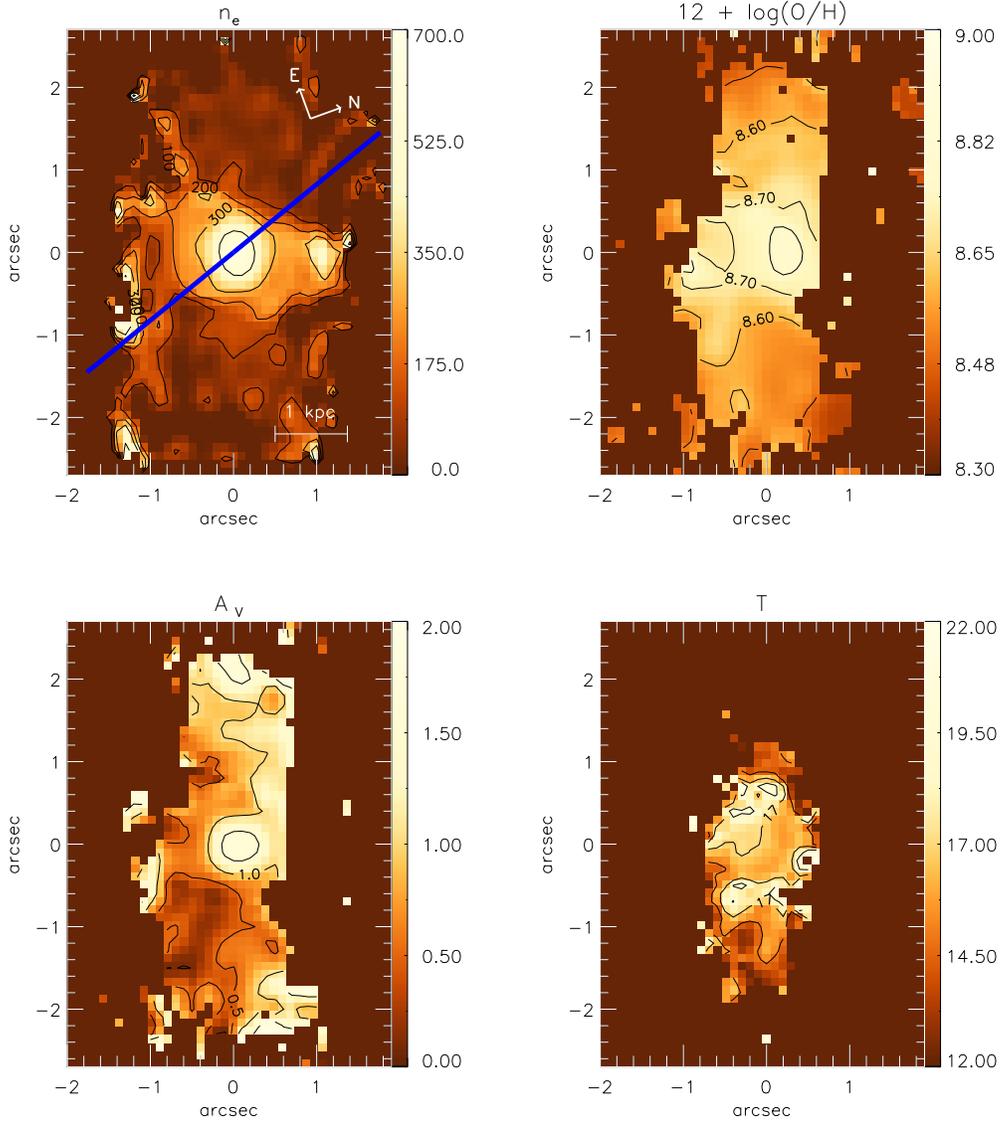}
\caption{From upper left to bottom right: electron density, oxygen abundance, visual extinction distribution maps and gas temperature. Density units are \cm, $A(V)$ is in magnitudes and temperature is in $\times 10^3\,$K. The blue line displays the radio jet axis.}
\label{dens_metal}
\end{figure*}

The [S\,{\sc ii}] ratio map allows us to derive the electron density distribution \citep{osterbrock06}, which is shown in the upper left panel of Fig. \ref{dens_metal}. The adopted temperature was $10,000\,$K. The highest density values ($> 500\,$\cm) are observed at the nucleus, with intermediate values of $\sim 200\,$\cm along the nuclear strip, with lower values in the range $50$-$100\,$\cm in the extended emission to the north-east and south-west. This is in agreement with the scenario in which the nuclear strip is formed by the passage of the radio jet, due to compression of the surrounding gas, which is thus at least partially excited by shocks.

The upper right panel of Fig. \ref{dens_metal} shows the oxygen abundance (O/H) distribution for the emitting gas, derived from the relations of \citet{storchi98}. As expected, the metallicity is higher in the nucleus, with values of $12 + \log(\rm{O/H})\sim 8.7$. A slightly higher metallicity is observed $\approx 0\farcs3$ north-west from the nucleus. Although 3C\,33 shows metallicity values typical of AGNs, these values are somewhat lower than those observed for Seyfert galaxies (typically $8.9 - 9.3$), and is actually comparable to values displayed by LINERs \citep{storchi98}. This suggests that the origin of part of the ionized gas may be the capture of a gas rich, lower metalicity galaxy, as is probably the case of another known radio galaxy, Pictor\,A, for which we performed a similar study in \citet{couto16}.

The H$\alpha$/H$\beta$ ratio map of Fig. \ref{ghratios} was used to obtain the visual extinction $A_V$ map. We adopted the reddening law from \citet{cardelli89}, and assuming case B recombination \citep{osterbrock06} we obtain

\begin{equation}
A_V = R_V\,E(B-V) = 6.9 \times \log \left( {\frac{H\alpha/H\beta}{3.1}} \right)
\end{equation} 

The corresponding visual extinction map is shown in the bottom left panel of Fig. \ref{dens_metal}. The highest values ($\sim 1.5$ mag) are observed at the nucleus, while values of $\sim 1$ mag are observed $\sim 1''-2''$ north-west of the nucleus, in the region where a faint dark lane is observed in the optical continuum acquisition image (see top central panel of Fig. \ref{large}). We propose that this lane is located along an inclined disk that we observe in rotation in the gas velocity map, as discussed below.

Finally, the bottom right panel of Fig. \ref{dens_metal} displays the temperature of the gas, as derived from the [O\,{\sc iii}]$\lambda$4959+5007/4363 ratio. We also used the [S\,{\sc ii}] ratio map as input parameter in the {\sc PyNeb} code \citep{luridiana15}, to derive the temperature map (which is shown in units of $\times 10^3\,$K). Although the coverage of the region where we could measure the temperature is more limited than other line ratio maps, we find the highest temperatures of about $\sim 18000\,$K along the nuclear strip, in particular in the regions where we observe the highest velocity dispersion values. The fact that the temperature increases outwards from the nucleus to drop again at farther distances indicate that the dominant mechanism responsible for this increase is not photoionisation by the AGN, but shocks. This is also supported by the higher velocity dispersion and line ratios values (e.g. [N\,{\sc ii}]/H$\alpha$), as we previously pointed out. In the following secion we discuss further evidence for the presence of shocks.

\subsection{Gas rotation}
\label{rot_model}

\begin{figure*}
\centering
\includegraphics[width=\textwidth]{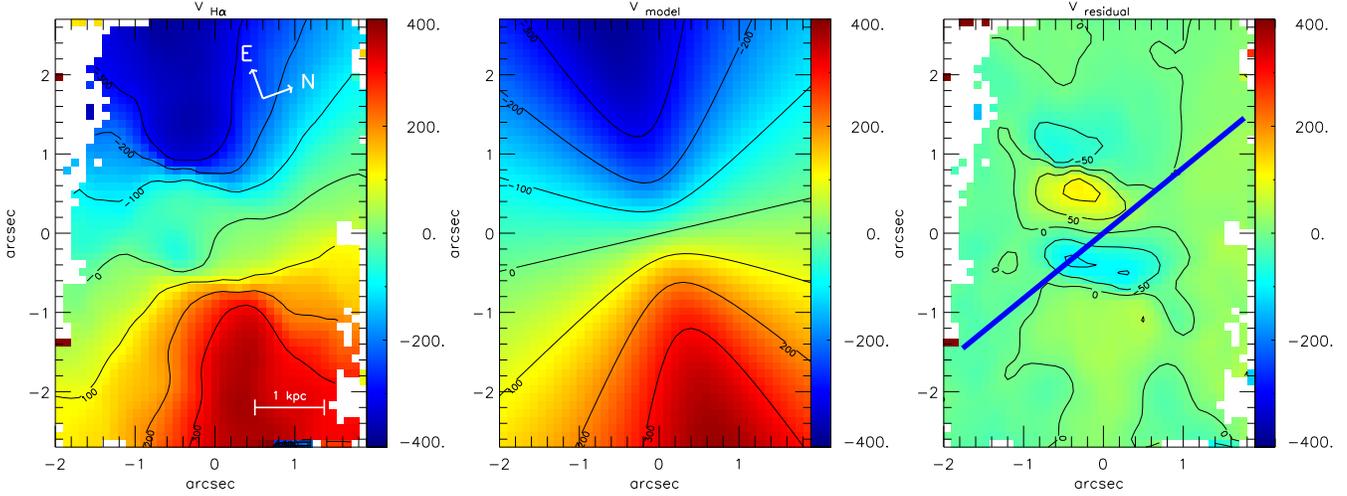}
\caption{Gaseous rotating disk model for H$\alpha$ emission line. Left: H$\alpha$ velocity map obtained from the Gauss-Hermite fit. Central: model of the rotation curve. Right: residual map. The blue line displays the radio jet axis.}
\label{rotmodel}
\end{figure*}

Centroid velocity maps of the ionized gas display a typical rotation pattern, which is traced both by the Gauss-Hermite fit in Fig. \ref{ghmaps} and by the narrow-component Gaussian fit in Fig. \ref{2gfit}. We fitted a kinematic model of circular orbits in a plane \citep{vanderkruit78,bertola91} to the H$\alpha$ centroid velocity field derived from the Gauss-Hermite fit. The model is described by the equation:

\begin{equation}
\begin{split}
v&_{mod} (R, \Psi) = v_{sys} + \\
&{\frac{A R\, cos (\Psi - \Psi_0)\, sin \theta\, cos^p \theta}{\{R^2[sin^2(\Psi - \Psi_0) + cos^2 \theta\, cos^2 (\Psi - \Psi_0)] + c_0^2\, cos^2 \theta\}^{p/2}}}
\end{split}
\label{vmode}
\end{equation}

\noindent
where $v_{sys}$ is the systemic velocity, $A$ is the centroid velocity amplitude, $r$ and $\Psi$ are the radial and angular coordinates of a given pixel in the plane of the sky, $\Psi_0$ is the position angle of the line of nodes, $c_0$ is a concentration parameter (constraining the radius at which the centroid velocity reaches $70$\% of the amplitude $A$) and $\theta$ is the disk inclination ($\theta = 0$ for a face-on disk). Finally, the parameter $p$ measures the slope of the rotation curve after reaching the maximum amplitude. This parameter was fixed at $p = 1$, which corresponds to an asymptotically flat rotation curve at large radii. The model also returns the center of rotation $(x_0,y_0)$ in terms of the distance to a reference pixel, which in our case we consider to be the peak of the continuum.

We used a Levenberg-Marquardt least-squares algorithm to obtain the best-fit model shown in Fig. \ref{rotmodel}. The resulting parameters from the fit are displayed in Table \ref{param}. Uncertainties are smaller than 5\% for all parameters.

The rotation model velocity map (central panel of Fig. \ref{rotmodel}) is a very good representation of the H$\alpha$ velocity field, confirming that it is dominated by rotation. The residuals, shown in the right panel of Fig. \ref{rotmodel}, are small and very close to zero, except for two regions with residuals of $\sim 100\,$\kms, one $\sim 0\farcs5$ east from the nucleus showing redshifts and another $\sim 0\farcs5$ west showing blueshifts. These residuals are in agreement with the scenario we propose in which this region is tracing a lateral expansion of the gas, pushed away by the radio jet. We also note that these regions are where we observe the highest [N\,{\sc ii}]/H$\alpha$ values in the channel maps of Fig. \ref{cmNIIHa}. As the near side of the galaxy is probably the north/north-west, where we observe the dust lane, a possible interpretation for the above residuals are the result of the lateral expansion of the gas by the shock produced by the ratio jet, with blueshifted residuals in the near side of the disk and redshifts in the far side of the disk. Another region, located $\sim 1\farcs2$ east from the nucleus, present substantial residual values of $\sim -50\,$\kms. However, we note that these residuals are an artificial feature generated from the higher residuals located close to it, along the nuclear strip. We re-fitted the rotation model to the velocity map, masking out the nuclear strip, and we obtained residuals close to zero in this region.

The best-fit rotating disk parameters give an inclination of $\theta = 66^\circ$, with velocity amplitude of $A = 453.7\,$\kms. This amplitude is very high for rotating disks originated from secular motions within the galaxy \footnote{see \citet{sofue99}, e.g., where the authors found that spiral galaxies usually present rotating velocities of $200-250\,$\kms at $\sim 2\,$kpc radius.}, and suggests that the rotating gas was recently acquired via an interaction, a process quite common in radio galaxies \citep{ramos11}. The orientation of the line of nodes, $\Psi_0 = 83.3^\circ$, is tilted by $\sim 10^\circ$ relative to the ionization axis (PA $\sim 70^\circ$). This means that the ionization cone, whose axis probably coincides with that of the radio jet, is intercepting a narrow patch of the disk. This is supported by the [O\,{\sc iii}] channel maps (Fig. \ref{cmOIII}) that show that the bulk of the emission is not observed at the highest velocities (along the major axis, where velocities of up to $\sim \pm 400\,$\kms are observed) but at velocities $\sim \pm 300\,$\kms.

\begin{table}
   \centering   
   \caption{\it Best-fit parameters from the disk kinematic model.}
   \begin{tabular}{|c|c|c|c|c|c|c|c|c|c|} 
      \hline \hline
      $\Psi_0$ (degree) & $83.3$  \\
      $\theta$ (degree) & $65.7$  \\
      $c_0$ (arcsec) & $1.2$  \\
      $A$ (\kms) & $453.7$  \\
      $v_{sys}$ (\kms) & $17801.8$  \\
      $x_0$ (arcsec) & $0.06$  \\
      $y_0$ (arcsec) & $0.01$ \\
      $p$ [fixed] & $1$  \\
      \hline
   \end{tabular}
   \label{param}
\end{table}

\begin{figure*}
\centering
\includegraphics[width=\textwidth]{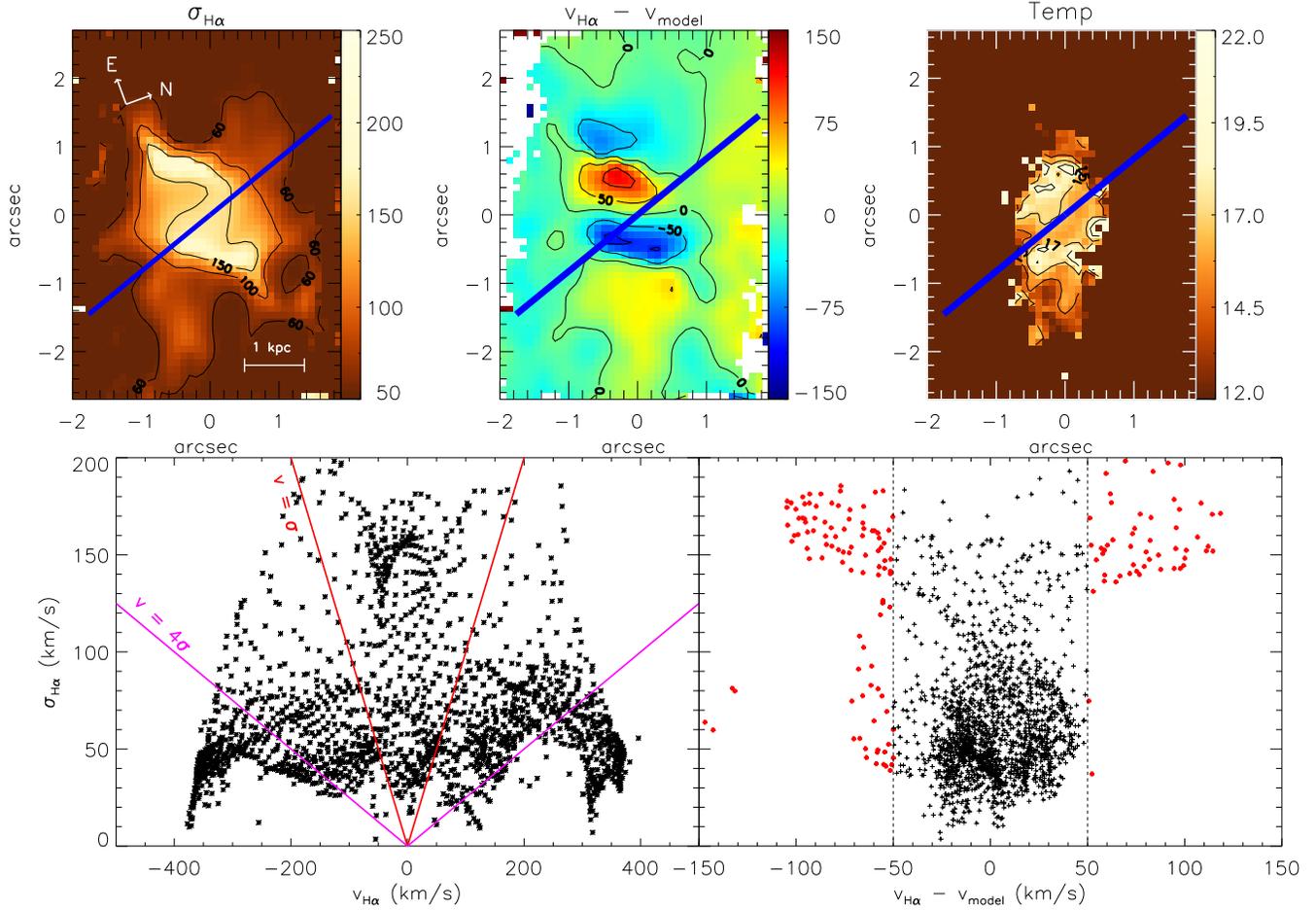}
\caption{Top panels: H$\alpha$ velocity dispersion, rotation model residuals and gas temperatures, as already shown in Figs. \ref{ghmaps}, \ref{rotmodel} and \ref{dens_metal}. The blue line displays the radio jet axis. Velocity and temperature units are \kms and $\times 10^3\,$K, respectively. Bottom panels: Diagrams of the relations of the H$\alpha$ velocity dispersion with H$\alpha$ centroid velocity (left) and rotation model residuals (right). Each asterisk represents a pixel of the GMOS FoV. The red and magenta lines in the left diagram represent $v = \sigma$ and $v = 4\sigma$, respectively. The red asterisks in the right diagram represent pixels where the residuals are greater than $\pm50\,$\kms, delimited by the dashed line.}
\label{velsig}
\end{figure*} 

\subsection{The nuclear strip}

Fig. \ref{velsig} illustrates the signature of shocks in the gaseous kinematics, in particular along the nuclear strip. The upper panels display the gas velocity dispersion, the residuals between the gas velocities and the rotation model and the gas temperature side by side to show that the residuals are highest ($> \pm50\,$\kms) in the regions where the velocity dispersion values are also the highest ($> 150\,$\kms), accompanied with an increase of the temperature that reaches $\sim 18000\,$K in these regions. In the bottom panels we display the relation between the H$\alpha$ velocity dispersion and both the H$\alpha$ centroid velocity and model residuals, for each pixel of the GMOS FoV. In the $\sigma_{H\alpha}$ vs. $v_{H\alpha}$ diagram, we can distinguish two main kinematical regimes: (1) pixels tracing the rotation present low velocity dispersion ($\sim 50\,$\kms) and are found closer to the $v = 4\sigma$ line; (2) pixels tracing the shocked and expanding gas along the nuclear strip present higher velocity dispersion ($> 100\,$\kms) and are found closer to the $v = \sigma$ relation. In the $\sigma_{H\alpha}$ vs. residuals diagram, pixels with low velocity dispersion have velocity residuals close to zero, while pixels with high velocity dispersion have higher velocity residuals. These diagrams present strong evidence of outflowing gas in the two regions of enhanced velocity dispersion. Added with the observed increase of the line ratios, electronic density and gas temperature, they support our interpretation that the outflow is producing shock ionization of the gas.

\subsubsection{Mass outflow rate and outflow kinetic power}

In this section we estimate the mass outflow rate related to the nuclear strip. Analyzing the outflow shape in the residuals distribution as seen in Fig. \ref{velsig}, we assume that the outflow is located along a inner radius of the rotating disk, in a section of it. We thus assume a cylindrical geometry, with disk-like dimensions, with a base radius extending up to the edge of the observed residual centered at the nucleus, $r = 0\farcs8$, and a small height of $h = 0\farcs2$. The mass flowing through the cross section of the assumed geometry is obtained from:

\begin{equation}
\dot{M}_{out} = n_e\,m_p\,v_{out}\,A\,f\, ,
\label{mout}
\end{equation}

\noindent
where $m_p = 1.7 \times 10^{-24}\,$g is the proton mass, $n_e$ is the electron density, $v_{out}$ is the velocity of the outflow perpendicular to $A = 2\pi\,r\,h = 1.3 \times 10^{43}\,$cm$^2$, which is the cross section of the disk edge, and $f$ is the filling factor.

We may estimate the filling factor using eq. \ref{eq2}. The volume of the assumed geometry is $V = 1.8 \times 10^{64}\,$cm$^3$, the mean electron density and the H$\alpha$ luminosity in the nuclear strip is $\approx 300\,$\cm and $2.6 \times 10^{41}\,$\ergs, respectively, which results in $f = 1.3 \times 10^{-4}$. Assuming the outflow velocity to be the highest velocity observed in the residuals, $v_{out} = 118\,$\kms, we obtain from eq. \ref{mout} $\dot{M}_{out} = 0.15\,$\msunyr. However, we are probably underestimating the outflow velocity, since it must have a projection angle with the line of sight. We may assume that the bulk of the outflowing gas is located along the rotating disk modelled in Sec. \ref{rot_model}. Thus, correcting the outflow velocity by a projected angle of $24.3^\circ$, we obtain that $v_{out} = 129\,$\kms, resulting in an outflow mass rate of $\dot{M}_{out} = 0.17\,$\msunyr.

We now use these estimations to calculate the outflow kinetic power \citep{holt06}, as

\begin{equation}
\dot{E} \approx {\frac{\dot{M}_{out}}{2}} (v_{out}^2 + 3\sigma^2) ,
\label{kinetic}
\end{equation}

\noindent
with the mean H$\alpha$ velocity dispersion $\sigma = 166\,$\kms at the nuclear strip. Using both the observed and projected outflow velocities, we obtain a range $4.7 < \dot{E} < 5.3 \times 10^{39}\,$\ergs.

We can also compare these values with the AGN bolometric luminosity. We estimate that $L_{bol} \approx 100 \times L(H\alpha)$, where $L(H\alpha)$ is the total observed H$\alpha$ luminosity, as we calculate in Sec. \ref{gas_mass}. We estimate that the AGN bolometric luminosity for 3C\,33 is $L_{bol} = 1.7 \times 10^{44}\,$\ergs, implying that the outflow kinetic power is only $2.6 < \dot{E}/L_{bol} < 3.0 \times 10^{-3} \%$. These values are comparable to estimates of outflow kinetic power of ionized gas in other local radio galaxies, such as Arp\,102B and 3C\,293 \citep{couto13,mahony16}. However, the bulk of outflowing gas mass in these objects is usually found in neutral and molecular phases \citep[i.e.][]{nesvadba10,feruglio10,garcia_burillo14}, and this is probably the case also for 3C\,33. In these phases, mass outflow rates are up to 2-3 orders of magnitude higher than for ionized gas. Also, we note that, considering this is a FRII radio galaxy presenting a large-scaled jet, most of the feedback probably occurs in the surrounding intergalactic medium.

\subsection{Emitting gas mass}
\label{gas_mass}

\begin{figure}
\centering
\includegraphics[width=0.5\textwidth]{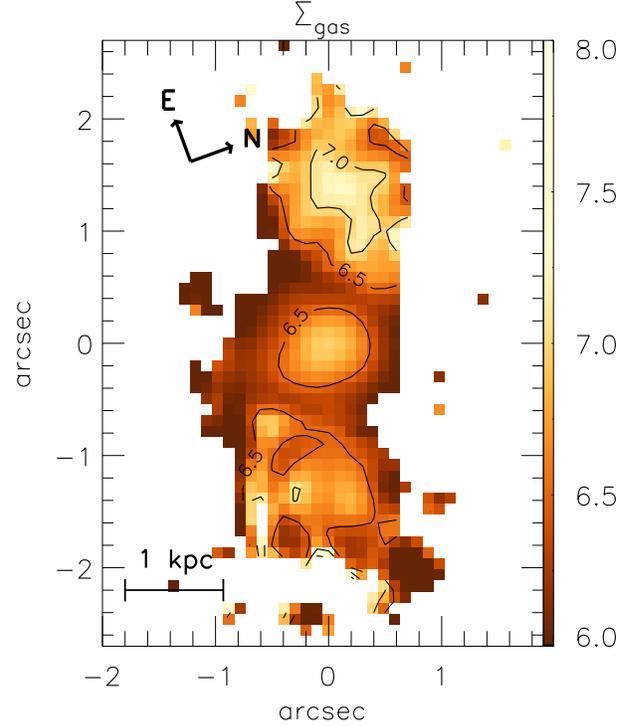}
\caption{Ionized gas mass distribution of 3C\,33. Units are M$_\odot$ and are in logarithmic scale. Regions where H$\beta$ were masked out due to high uncertainties were also masked out in this map. However the mass values decrease to negligible values in these regions, and does not affect the total mass estimative.}
\label{gasmassfig}
\end{figure}

We can estimate the mass of the ionized gas as \citep{peterson97}:

\begin{equation}
M \approx 2.3 \times 10^5\,{\frac{L_{41}(\textrm{H}\alpha)}{n_3^2}}\,\textrm{M}_\odot\, .
\end{equation}

\noindent
where $L_{41}(\textrm{H}\alpha)$ is the H$\alpha$ luminosity in units of $10^{41}\,$erg s$^{-1}$ and $n_3$ is the electron density in units of $10^3$\,\cm.  

We corrected the total H$\alpha$ luminosity for reddening, assuming $R_V = 3.1$ from \citet{cardelli89}, resulting in $L(\textrm{H}\alpha) = 4\pi d^2\,F(\textrm{H}\alpha)\,10^{C(\textrm{H}\alpha)} = 1.7 \pm 0.1 \times 10^{42}\,$\ergs. Using the electron density map shown in Fig. \ref{dens_metal}, we could map the ionized gas mass distribution over our FoV, as shown in Fig. \ref{gasmassfig}. The total ionized gas mass within the inner $\sim 2.5\,$kpc radius results in $M = 4.1 \pm 1.7 \times 10^8\,$M$_\odot$. This mass is consistent with the values obtained in previous studies of other radio galaxies, such as that of \citet{tadhunter14}, who estimate that strong-line radio galaxies (like 3C\,33) with redshifts between $0.05$ and $0.7$ have gas masses of $1.0 \times 10^8 < M < 3.7 \times 10^{10}\,$M$_\odot$, with a median of $1.2 \times 10^9\,$M$_\odot$.  

\section{Conclusions}
\label{conc}

We have measured the gaseous kinematics and excitation in the inner $4.0\,$kpc\,$\times 5.8\,$kpc of the narrow line radio galaxy 3C\,33, from optical spectra obtained with the GMOS integral field spectrograph on the Gemini North telescope, at a spatial resolution of $\approx 580\,$pc at the galaxy. 

We observe elongated gas emission extending to $\approx 2\,$kpc north-east and south-west of the nucleus, a direction that we identify as that of the ionization axis. We could separate two kinematical components in the GMOS FoV: a broader component, with velocity dispersions of $\sigma \ge 150\,$\kms, which is dominant within a $1\,$kpc wide ``nuclear strip'' -- running from south-east to the north-west, approximately perpendicular to the direction of the radio jet -- and a narrower component, with velocity dispersions of $\sigma \le 100\,$\kms, which dominates the emission beyond the nuclear strip.

The narrow component traces the rotation pattern of the gas, observed in the centroid celocity maps. We obtain a good fit when modeling the H$\alpha$ velocity field with a rotation model with a velocity amplitude of $\sim 450\,$\kms and an angle between the disk plane and the plane of the sky of $\sim 65^\circ$. Residual blueshifts and redshifts of $\sim 100\,$\kms are observed within the nuclear strip. Rotation is also observed in the channel maps of the [O\,{\sc iii}]$\lambda$5007 emission line, in which we observe that the bulk of emission is located in velocity bins close to $\sim 300\,$\kms, corresponding to the orientation of the ionization axis, while the emission at higher velocity bins -- and lower intensities -- are shifted to a higher position angle, in agreement with the orientation of the line of nodes of the disk model ($65^\circ$). Regions of highest emission are also extended along the radio jet axis in the velocity channels close to zero (in the galaxy velocity frame).

The nuclear strip is characterized by high velocity dispersions ($\sim 170\,$\kms), highest [N\,{\sc ii}]/H$\alpha$, [S\,{\sc ii}]/H$\alpha$ and [O\,{\sc i}]/H$\alpha$ line ratios, highest velocity residuals from the rotation model, high electron density ($> 300\,$\cm) and highest gas temperatures ($\sim 18000\,$K). We conclude that this region is tracing a gas outflow in lateral expansion, due to the passage of the radio jet, which is oriented approximately perpendicularly to the nuclear strip. The caracteristics of this region all point to shocks being the main mechanism of gas excitation. Photoionization from the AGN radiation seems to dominate along the ionization axis and the jet axis, where we have observe increased values of [O\,{\sc iii}]/H$\beta$ line ratio.

We estimate small values of mass outflow rate and outflow kinetic power, $0.15 < \dot{M}_{out} < 0.17\,$\msunyr and $4.7 < \dot{E} < 5.3 \times 10^{39}\,$\ergs, respectively. This represents only $2.6 < \dot{E}/L_{bol} < 3.0 \times 10^{-3}\,$\%. These values are comparable to estimates of kinetic power related to warm outflowing gas in other local radio galaxies. However, in order to better understand the properties of the outflow, further observations of molecular gas emission would be necessary.

Although the gas kinematics seems to be dominated by rotation (in the disk) and outflow (along the nuclear strip), we observe a possible signature of gas inflow: increased [O\,{\sc i}]/H$\alpha$ values along two spiral arms in the disk, where we also observe an increase in velocity dispersion. We interpret that these arms trace shocks in the rotating gas, which allows a decrease in angular momentum resulting in gas inflow.

We have obtained a total mas of ionized gas within the inner $\sim 2.5\,$kpc of $M = 4.1 \pm 1.7 \times 10^8\,$M$_\odot$. Its high rotation velocity amplitude ($\sim \pm 450\,$\kms) and the somewhat lower metallicity ($12 + \log \rm{(O/H)} \sim 8.5 - 8.8$) than typically observed in the nuclear region of AGN suggests that this gas has been recently acquired, in an interaction event that was probably the trigger of the nuclear activity in 3C33.

\section*{Acknowledgments}

This work is based on observations obtained at the Gemini Observatory, which is operated by the Association of Universities for Research in Astronomy, Inc., under a cooperative agreement with the NSF on behalf of the Gemini partnership: the National Science Foundation (United States), the National Research Council (Canada), CONICYT (Chile), the Australian Research Council (Australia), Minist\'{e}rio da Ci\^{e}ncia, Tecnologia e Inova\c{c}\~{a}o (Brazil) and Ministerio de Ciencia, Tecnolog\'{i}a e Innovaci\'{o}n Productiva (Argentina). This work has been partially supported by the Brazilian institutions CNPq and FAPERGS.

\bibliographystyle{mn2e}
\bibliography{refs}

\end{document}